\documentclass[floats,floatfix,showpacs,amssymb,prd,twocolumn,superscriptaddress,nofootinbib,reprint]{revtex4-2}

\usepackage{amssymb,amsmath,verbatim,mathtools,needspace,enumitem,etoolbox,graphicx,physics,microtype,afterpage,xspace,tabularx,lmodern,multirow}
\usepackage{siunitx}
\usepackage{gensymb}
\usepackage{ulem}  
\usepackage[dvipsnames]{xcolor}
\definecolor{linkcolor}{rgb}{0.0,0.3,0.5}
\usepackage[unicode, colorlinks=true, linkcolor=linkcolor, citecolor=linkcolor, filecolor=linkcolor, urlcolor=linkcolor, linktocpage, breaklinks]{hyperref}
\usepackage[all]{hypcap}
\usepackage[T1]{fontenc}
\usepackage[utf8]{inputenc}
\hypersetup{colorlinks=true,citecolor=romared,linkcolor=romared,urlcolor=romared}

\definecolor{romared}{RGB}{142,0,28}

\def\be{\begin{equation}}
\def\ee{\end{equation}}

\usepackage{aas_macros}
\usepackage{makecell}
\usepackage{soul}

\usepackage{lipsum}
\usepackage{orcidlink}

\newcolumntype{Y}{>{\centering\arraybackslash}X}

\begin{document}
\title{The Rise and Fall of Acoustic Oscillations at Cosmic Dawn}

\author{Hector Afonso G. Cruz\,\orcidlink{0000-0002-1775-3602}}
\email{h.cruz@nyu.edu}
\affiliation{Center for Cosmology and Particle Physics, Department of Physics, New York University, New York, NY 10003, USA}
\affiliation{William H. Miller III Department of Physics and Astronomy, Johns Hopkins University, 3400 N. Charles Street, Baltimore, Maryland, 21218, USA}

\author{Gabriele Montefalcone\,\orcidlink{0000-0002-6794-9064}}
\affiliation{Texas Center for Cosmology and Astroparticle Physics,
Weinberg Institute for Theoretical Physics,
University of Texas at Austin, Austin, TX 78712, USA}

\author{Julian B. Mu\~noz\,\orcidlink{0000-0002-8984-0465}} 
\affiliation{Department of Astronomy, The University of Texas at Austin, 2515 Speedway, Stop C1400, Austin, Texas 78712, USA}
\affiliation{Cosmic Frontier Center,
University of Texas at Austin, Austin, TX 78712, USA}
\affiliation{Texas Center for Cosmology and Astroparticle Physics,
Weinberg Institute for Theoretical Physics,
University of Texas at Austin, Austin, TX 78712, USA}

\author{Ely D. Kovetz\,\orcidlink{0000-0001-9256-1144}} 
\affiliation{Department of Physics, Ben-Gurion University of the Negev, Be’er Sheva 84105, Israel}

\author{Alessandra Venditti\,\orcidlink{0000-0003-2237-0777}}
\altaffiliation{Cosmic Frontier Center Prize Fellow}
\affiliation{Department of Astronomy, The University of Texas at Austin, 2515 Speedway, Stop C1400, Austin, Texas 78712, USA}
\affiliation{Cosmic Frontier Center,
University of Texas at Austin, Austin, TX 78712, USA}

\begin{abstract}
Cosmic dawn 21-cm observations will extend standard-ruler cosmology into the first billion years, unlocking epochs inaccessible by the cosmic microwave background and large-scale structure. Realizing this promise requires an accurate model of the acoustic structure imprinted onto early star formation. At such early times, two counterbalancing phenomena---matter overdensities and streaming velocities between cold dark matter and baryons---modulate the spatial statistics of star formation. While overdensities dictate where early galaxy-bearing haloes form, regions of high velocity suppress star formation. Their combined influence on the intergalactic medium yields 21-cm fluctuations with both baryon (BAO) and velocity-induced (VAO) acoustic oscillations. Here we present the first prescription to decompose the 21-cm power spectrum into its constituent acoustic features. We find a percent-level offset between the BAO and VAO shapes which, if ignored in standard-ruler analyses, would bias inferred values of $H(z)$ by $\sim 2\%$; we provide a correction. Moreover, as the relative prominence of BAOs and VAOs ebbs and flows non-monotonically across cosmic dawn, we demonstrate how their evolution is sensitive to the physics of early galaxy evolution and the first stars. Finally, we forecast how sensitive SKA will be to the BAO-VAO combined standard ruler. Our results establish joint BAO-VAO modeling as an essential ingredient of 21-cm acoustic inference, enabling robust constraints on both cosmic expansion and the first stars.
\end{abstract}

\date{\today}
\maketitle

\textbf{Introduction---} Acoustic oscillations have become one of the principal tools for measuring cosmological parameters. 21-cm line intensity maps from neutral hydrogen can extend their constraining potential to unprecedentedly higher redshifts, when the first luminous sources transformed the intergalactic medium (IGM). In the era before observable structure, overdense regions collapsed to form nascent dark matter (DM) haloes, spawning the first generation of luminous sources at cosmic dawn (redshifts $15 \lesssim z \lesssim 35)$ \citep{bromm13, klessen23}. These stars pervade the cosmos with ultraviolet (UV) and X-ray photons, whose heating, ionizing, and coupling effects on the IGM yield an initial epoch of absorption \citep{wouthuysen52, field59, loeb04, hirata06} and subsequent emission \citep{Furlanetto:2006jb, pritchard12, ciardi10, xu14, ewall16, sazonov17} until the conclusion of reionization (at $z \sim 5)$. Such early times are difficult to explore with traditional surveys as the first galaxies are too far, faint, and redshifted. Thus, current and upcoming  interferometers~\cite{vanhaarlem13, mellema13, voytek14, bowman18, beardsley16, greig21, mertens20, yoshiura21} (including the \textit{Square Kilometer Array} (SKA) \citep{carilli04, furlanetto04b, murphy09, santos11, koopmans15, ghara16, barry16, hutter17, hassan20}) will yield insight into the elusive astrophysics of the first galaxies and their cosmological initial conditions \citep{loeb01, Barkana:2004zy, Furlanetto:2006tf}.

Two competing ingredients rooted in pre-recombination physics dictate which haloes first form stars. Although DM began to cluster gravitationally, radiation pressures prevented baryons from following suit. This created the well-known baryon acoustic oscillation (BAO) feature \citep{Peebles:1970ag, Sunyaev:1970bma} used as a standard ruler in cosmic microwave background (CMB) \citep{WMAP:2003elm, WMAP:2012nax, planck18, ACT:2025fju} and large-scale structure (LSS) \citep{eisenstein98, SDSS:2005xqv, eBOSS:2020yzd, DES:2024pwq, DESI:2025zgx} analyses. Simultaneously, these radiation-driven density waves induced a supersonic relative velocity between baryons and cold DM \citep{tseliakhovich10}. At $z_\mathrm{kin} \approx 1060$, the universe cools sufficiently such that photons kinematically decouple from baryons, imprinting both a matter density ($\delta(\textbf{x})$) and a velocity differential ($\eta(\textbf{x}) \equiv v^2_\mathrm{cb}(\textbf{x},z)/\sigma_\mathrm{cb}^2(z) - 1$) field encoded with the acoustic scale from their inception, where $v^2_\mathrm{cb}$ is the squared velocity magnitude. The latter impedes later star formation by suppressing small-scale matter fluctuations \citep{tseliakhovich10, naoz12, bovy13}, disrupting the cooling efficiency of protogalactic gas cores \citep{greif11, fialkov12, hirano18, schauer19}, and inhibiting accretion onto early haloes \citep{dalal10, tseliakhovich11, stacy11, oleary12, naoz13}.

Born in overdense regions and muted by relative velocities, the first star-forming galaxies (SFGs) eventually heat and ionize their neutral surroundings. Early SFGs illuminated the IGM and produced a 21-cm power spectrum with superimposed BAOs and ``velocity-induced acoustic oscillations'' (VAOs). Seeded only by cosmology, the positions of VAO extrema are robust to astrophysics \citep{munoz19}. However, a host of early-universe phenomena offset the phase between BAO and VAO wiggles in 21-cm correlations \citep{Yoo:2011tq, visbal12, alihaimoud14, ferraro12,  Beutler:2016zat, Schmidt:2016coo, Schmidt:2017lqe, Givans:2020sez}. Consequently, the 21-cm acoustic signal is not simply a high-redshift copy of low-redshift BAOs, but a mixed acoustic ruler whose components trace different physics. This distinction is crucial: both acoustic peaks are standard rulers that can be used to measure cosmic expansion \citep{munoz19b, Sarkar:2022mdz}, while their amplitudes can probe small-scale structure \citep{munoz20, Zhang:2024pwv, verwohlt2024}, early galaxy formation \citep{Libanore:2023oxf}, DM models \citep{Hotinli:2021vxg}, and other exotic physics \citep{Hotinli:2021xln, cruz24, Montefalcone:2025mbg}.

IGM acoustic signatures can therefore probe completely uncharted epochs, and will place complementary constraints to those imposed by the early CMB and later LSS. But the key to percent-level inference on either astrophysics or cosmology relies on understanding how BAOs and VAOs add, compete with, and differ from each other over time and scale. Though BAO statistics show that densities fluctuate at the percent level from the background \citep{eisenstein98, ma95}, velocities fluctuate at order unity \citep{tseliakhovich10, tseliakhovich11}. Additionally, velocities decay non-uniformly over redshift \citep{hahn21, ferraro12}, and suppress star formation in only the lowest-mass haloes \citep{mcquinn12, schauer19, kulkarni21}. All these effects imply that VAOs may dominate the 21-cm signal over BAOs toward higher redshifts before haloes accrete the bulk of their present-day mass. Lastly, VAOs briefly vanish in the transition between the Lyman-$\alpha$ coupling and X-ray heating epochs~\citep{Munoz:2020itp}, suggesting they also do so at the onset of reionization. Together, these puzzle pieces suggest that the rise and fall of either BAOs or VAOs may not always be monotonic or mutually distinguishable, and that a BAO-only treatment can lead to a systematically incorrect standard-ruler measurement.

In this \textit{Letter}, we study the competition between density and velocity fluctuations in the 21-cm signal at cosmic dawn, and show how significant detections of either ingredient may lead to discrepant measurements of cosmological parameters. Our dissection is only possible with the rapid fully-analytical public code {\tt\string Zeus21} \citep{munoz23, Cruz:2024fsv}. By decomposing the matter and relative velocity power spectra into oscillatory ``wiggles'' atop its baseline ``no-wiggle'' components, we derive BAO and VAO spectrum-based templates of 21-cm observables. After testing how either density or velocity can be confused in our spectral template decomposition, we forecast both detectability of, and distinguishability between, either acoustic features using SKA \citep{carilli04, furlanetto04b, murphy09, santos11, koopmans15, ghara16, barry16, hutter17, hassan20}. We find that joint BAO-VAO modeling is critical for 21-cm cosmology, without which its promising acoustic structure can become a source of systematic error. We use the \textit{Planck} 2018 \citep{planck18} cosmology and the Boltzmann solver {\tt\string CLASS} \citep{Blas:2011rf} throughout this work.


\begin{figure*}[t!]
    \centering
    \includegraphics[width=0.95\textwidth]{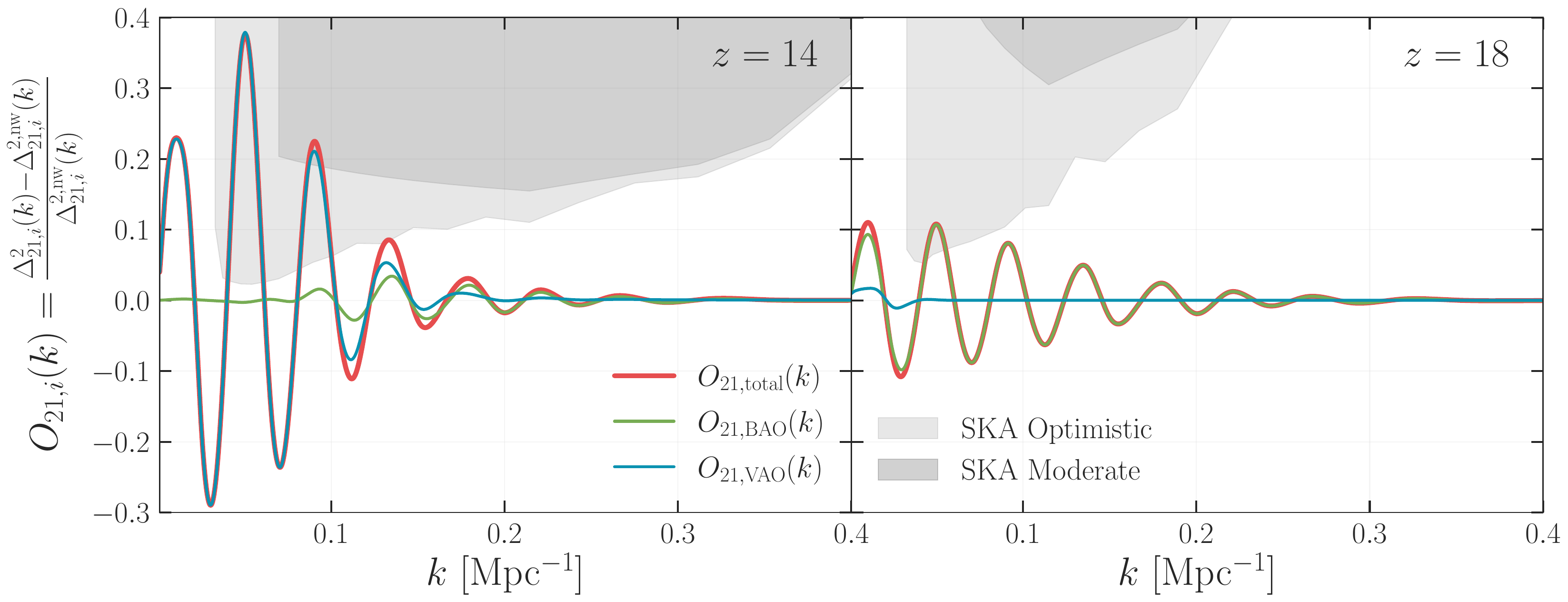}
    \caption{The normalized ``wiggle''-only portion of $\Delta^2_{21}(k,z)$ at two redshifts, split into its BAO (green), VAO (blue), and total (red) components. At $z\sim 14$ (left), VAOs dominate the first three peaks but quickly dampen toward higher $k$. The earlier $z_\mathrm{min} \sim 18$ (right, chosen at the transition between Lyman-$\alpha$ coupling and X-ray heating) case shows negligible VAOs, so BAOs dominate at all scales. Shaded regions correspond to SKA sensitivities assuming optimistic (light) and moderate (dark) foreground scenarios.
    }
  \label{fig:FIG1}
\end{figure*}



\textbf{Disentangling Acoustic Signatures---}\label{sec:disentanglingAcousticSignatures}
Here we describe our modeling of 21-cm fluctuations that can be bifurcated into its BAO and VAO components. The abundance of SFGs is enhanced in overdensities and attenuated in high-velocity regions, which modifies the local thermal and ionization history of the IGM \citep{visbal12, mcquinn12, Fialkov:2013uwm, munoz19} (see the Supplementary Material for astrophysical modeling details.) We use the {\tt\string Zeus21} framework to isolate how this density- and velocity-sourced acoustic structure propagates into the 21-cm power spectrum. Following Refs.~\citep{munoz23, Cruz:2024fsv}, fluctuations in the star formation rate density (SFRD) can be described as functionals of the density ($\delta$) and relative velocity ($\eta$) perturbations, which can be used as analytic building blocks to compute fluctuations in the Lyman-$\alpha$ coupling and X-ray heating terms in the 21-cm brightness temperature $T_{21}$. Using these $\delta$ and $\eta$ building blocks, {\tt\string Zeus21} computes the reduced power spectrum $\Delta^2_{21}(k,z) = k^3 P_{21}(k,z)/(2\pi^2)$ as sums of functionals of the matter power spectrum $P_m(k)$ and the scalar velocity spectrum $P_\eta(k)$. To isolate BAOs from VAOs in $\Delta^2_{21}(k,z)$, we treat their wiggles as small perturbations over the smooth baseline with mixed baryon/velocity cross-terms entering at vanishing second order. This allows us to isolate the oscillatory ``wiggle'' component from its smooth ``no-wiggle''(nw) baseline:
\begin{equation}
    \Delta^2_{21}(k,z) = \Delta^{2}_{21,\mathrm{nw}}(k,z)[1 + O_{21}(k,z)],
\end{equation} 
where the total oscillatory component $O_{21} = O_{21,\mathrm{BAO}}+O_{21,\mathrm{VAO}}$ is simply a sum of its matter and velocity ``wiggle'' components, normalized. We proceed by first isolating the ``wiggle'' and ``no-wiggle'' components from $P_m(k)$ and $P_\eta(k)$, then alternate these as inputs into {\tt\string Zeus21} to yield the 21-cm wiggle and no-wiggle spectra. Our baseline extraction procedure is outlined in the Supplementary Material. 

In Fig.~\ref{fig:FIG1}, we show the normalized ``wiggle'' component of the 21-cm power spectrum at two different times, decomposed into its BAO and VAO components. As explained by Ref.~\citep{Montefalcone:2025mbg}, one subtle characteristic across both times is that the BAO and VAO components are not perfectly in phase with each other. The first BAO bump is shifted slightly leftward of the VAOs, while the rest of the BAO peaks are shifted increasingly rightward of their corresponding VAO counterparts. Nevertheless, they add nearly coherently, producing a total acoustic feature that oscillates by up to $ 20-30\%$ from the baseline power spectrum across most times. At $z \sim 14$ toward the end of cosmic dawn, VAOs dominate at scales $k \lesssim 0.13 \, \mathrm{Mpc}^{-1}$ and decay with scale. BAOs overtake VAOs only at the fifth acoustic peak. At $z_\mathrm{min} \sim 18$ where the 21-cm global signal absorption trough is at its minimum, BAOs dominate all scales. 

Both snapshots can be explained by the competition between density- and velocity-driven astrophysical perturbations over time. The $T_{21}$ minimum marks the transition between the Lyman-$\alpha$ coupling and X-ray heating epochs. While $x_\alpha$ coupling forces a more negative signal, X-rays heat the IGM and drive a more positive $T_{21}$. High velocity regions produce less stellar photons, and will counteract the negative coupling in the former epoch and the positive heating in the latter. At the transition redshift close to $z_\mathrm{min} \sim 18$, velocity-modulated regions are trapped in this negative feedback loop that effectively erases $\eta$-induced anisotropies in the IGM, resulting in a temporary BAO-only signal. BAOs suffer the same regulation at the later redshift, as $x_\alpha$ coupling must counteract $\delta_b$-induced fluctuations in addition to X-rays. This phenomenon is counterintuitive; though VAOs should be most prominent at higher redshifts when haloes were less massive, this transitionary epoch is one critical exception.

\begin{figure}[t!]
    \centering
    \includegraphics[width=0.5\textwidth]{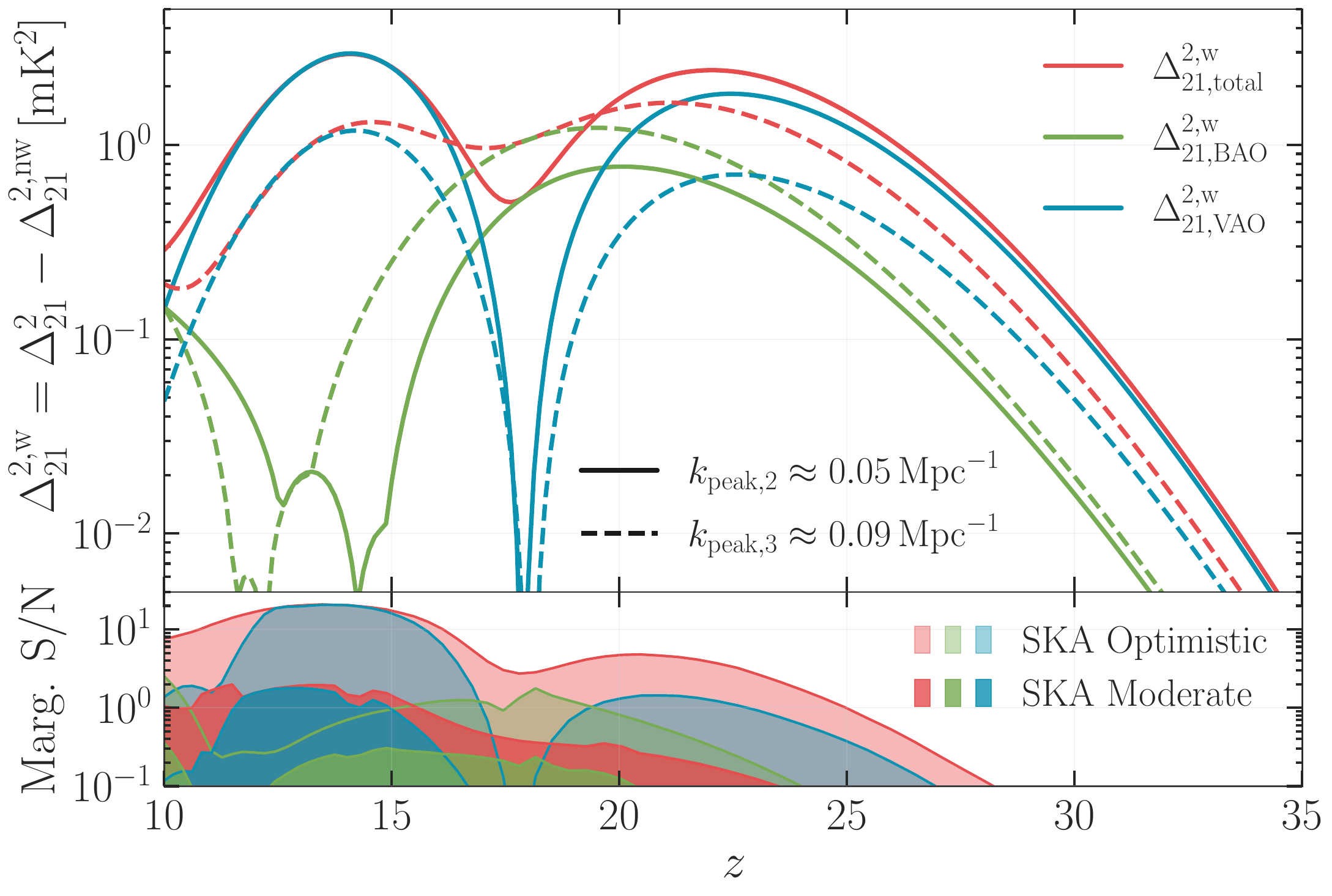}
    \caption{\textbf{(Top):} Amplitudes of the second (solid) and third (dashed) acoustic peaks for the total 21-cm spectrum (red) and its BAO (green) and VAO (blue) components. VAOs vanish at $z\sim 18$ where $T_{21}$ is at a global minimum, while BAOs vanish later at $z \sim 14$. \textbf{(Bottom):} Marginalized SNR forecasts for detecting acoustic oscillations with SKA. The ``moderate'' case (dark) can detect total oscillations at $13\lesssim z \lesssim 15$ by ($\sim 2 \sigma$), while the optimistic case (light) can detect them at $10\lesssim z \lesssim 24$. Since they add together nearly in phase, both acoustic signals boost the SNR; only in the high-SNR regime can they be disentangled from each other.}
  \label{fig:FIG2}
\end{figure}

Beyond the above two snapshots, BAOs and VAOs compete non-monotonically across redshift and scale in the 21-cm power spectrum. In Fig.~\ref{fig:FIG2}, we plot the amplitude of the second and third peaks of the unnormalized wiggle power spectrum $\Delta^{2}_{21, \mathrm{w}}(k,z) \equiv \Delta^{2}_{21, \mathrm{nw}}O_{21}$ over redshift, separated into its matter, relative velocity, and total components. The second VAO peak at $k\approx 0.05 \, \mathrm{Mpc}^{-1}$ is the tallest across all time, while the third BAO peak at $k\approx 0.09 \, \mathrm{Mpc}^{-1}$ is tallest at $z \gtrsim 13$. As expected, VAOs vanish around $z_\mathrm{min}$ while BAOs persist until $z \sim 14$. 

This behavior is explained by linearizing 21-cm fluctuations, 
    \begin{equation}
    \frac{\delta T_{21}(\mathbf{x}, z)}{\bar{T}_{21}(z)}=\beta_{b} \delta_b+\sum_{i \in (m, \eta)} \beta_\alpha \delta x_\alpha^{(i)}+\beta_X \delta T_X^{(i)}
    \end{equation}
whose weights $\beta$ are astrophysical biases defined in Ref.~\citep{Cruz:2024fsv}. While matter perturbations contribute to the LSS, Lyman-$\alpha$, and X-ray terms, relative velocities enter only through the radiation backgrounds. Around $z_\mathrm{min}\sim 18$, X-ray heating cancels out $x_\alpha$-coupling in velocity-modulated regions, suppressing VAOs in the 21-cm power spectrum. BAOs vanish later at $12 \lesssim z \lesssim 15$ as the Lyman-$\alpha$ intensity must grow to counteract both the X-ray and LSS terms. An auxiliary interpretation of the disappearance of either acoustic feature can be found in Ref.~\citep{Munoz:2020itp}. Since radiation affects the IGM from the inside-out, higher-$k$ bumps vanish before lower-$k$ bumps.


\textbf{Detectability vs. Separability---} \label{sec:DetectSeparate}
Measuring the temporal and spatial competition between BAOs and VAOs hinges on instrumental sensitivity and foreground avoidance. Though spectrally smooth foregrounds vary slowly along the line of sight, interferometers induce chromaticity that bleeds them across Fourier space termed the ``foreground wedge.'' Modes with wavenumbers below $k_{\|}^{\min }\lesssim a+b(z) k_{\perp},$ are contaminated, given a horizontal floor $a$ and slope $b$ \citep{parsons12, morales12, datta10, parsons12b}. 
We will assume two foreground removal scenarios, an ``optimistic'' case $(a=0, b(z)\approx 1)$ where the primary beam size defines the wedge, and a ``moderate'' case where the wedge extends beyond the horizon limit as $(a=0.05 \, h/\mathrm{Mpc}, b(z)\approx 6)$.
To forecast sensitivity and separability of both acoustic features, we use the public code {\tt\string 21cmSense} \citep{pober13, pober14} to simulate SKA-Low across $10 \lesssim z \lesssim 25$. To reach the largest scales, we use a ``HERA''-like drift scan survey including the 18 $\mathrm{m}$ substations in the AA4 assembly. We bin scales $0.02 \leq k \leq 0.4 \, \mathrm{Mpc^{-1}}$ logarithmically, which reduces sensitivity but better resolves the first three acoustic peaks. To mitigate the former, we simulate three one-year SKA observations at six hours per evening at 6, 7, and 8 $\mathrm{MHz}$ bandwidths, and co-add observations by adding inverse noise variances. 

In Fig.~\ref{fig:FIG1}, we plot the $1\sigma$ detectability region of both foreground scenarios rescaled by the ``no-wiggle'' baseline power spectrum as light and dark shaded gray bands. Across all times, BAO and VAO features add nearly coherently and are far more detectable than either alone. At the later $z\sim 14$ snapshot, the ``moderate'' case is just sensitive to the total third acoustic peak while the optimistic case encompasses the second and third VAO bumps. The earlier $z\sim 18$ snapshot shows that the ``optimistic'' scenario can detect only the second peak. At both times, the sensitivity region exhibits a sharp cut toward low $k$ from foreground avoidance and sample variance, and weakens toward higher $k$ from thermal noise.


Distinguishability between matter and velocity wiggles is a more stringent question than mere detectability. To that end, we study the relative prominence of either acoustic peaks over redshift, and perform an information matrix analysis \citep{Jungman:1995bz, Jimenez:2013mga, munoz20} by modeling the decomposed power spectrum as
\begin{equation}
    \Delta^2_{21} = A_\mathrm{nw}\Delta^{2, \mathrm{nw}}_{21} + A_\mathrm{BAO}\Delta^{2}_{21, \mathrm{BAO}} + A_\mathrm{VAO} \Delta^{2}_{21,\mathrm{VAO}}
\end{equation}
in terms of the ``no-wiggle'' baseline power spectrum $\Delta^{2, \mathrm{nw}}_{21}$ and the superimposed BAO ($\Delta^{2}_{21, \mathrm{BAO}}$) and VAO ($\Delta^{2}_{21,\mathrm{VAO}}$) wiggles, all implicitly functions of redshift and scale. Their amplitudes $A_i$ are fiducially unity, but are constructed as free amplitudes such that their information matrix at each redshift can be expressed with the {\tt\string 21cmSense}-derived sensitivities $\sigma^2(k)$ as
\begin{equation}
    F_{i j}(z)=\sum_k \frac{1}{\sigma^2(k)} \frac{\partial \Delta^2_{21}}{\partial A_i}(k) \frac{\partial \Delta^2_{21}}{\partial A_j}(k).
\end{equation}
The marginalized-SNR is simply $\mathrm{SNR}_i \equiv A_i/\sigma_i = 1 / \sqrt{(F^{-1})_{ii}}$, shown in the bottom panel of Fig.~\ref{fig:FIG2}. We focus only on the second and third peaks, as the first falls within the foreground-limited region and all other peaks are damped beyond observability.

Across all redshifts, VAOs overtake BAOs except for $17 \lesssim z \lesssim 19$ near the $T_{21}$ minimum. Using our stellar model, the second BAO peak maximum is half that of the VAO maximum, though the third peaks are equally prominent. While the second VAO peak is taller than the third, BAOs show the opposite until $z\lesssim 12$, suggesting that velocity-induced fluctuations have higher large-scale variance than matter-induced ones. BAO amplitudes jump toward lower redshifts, eventually overtaking VAOs by $z\lesssim 11$. The ``optimistic'' scenario shows SKA is sensitive to the total acoustic oscillation by $\gtrsim 2\sigma$ across the $10 \lesssim z \lesssim 24$ even with the dip around the $T_{21}$ minimum where VAOs briefly disappear. In this case, VAOs are most detectable around $12\lesssim z \lesssim 18$, even in the realm where VAO and BAO amplitudes are near parity right after VAOs vanish. The ``moderate'' case can only detect total acoustic oscillations by $\sim 2\sigma$ around $13 \lesssim z \lesssim 15$ where VAOs achieve their maximum amplitude. This epoch is also where VAOs are most likely to be detected as BAO amplitudes plummet. Across all foreground removal scenarios, BAOs are not significantly detected whereas VAOs can be at the right redshifts. Compared to HERA with a similar observational setup, SKA performs better on large-scale detectability by about $\sim 0.5 \sigma$ across ``moderate'' and by a factor of two across ``optimistic'' scenarios. In summary, ``optimistic'' scenarios are more sensitive to both total and VAO-only wiggles over a broader redshift range than ``moderate'' scenarios. Both acoustic wiggles together boost the detectability across a broader redshift support. Nonetheless, BAOs and VAOs may still be separated by the difference in their phases at redshifts when only the total wiggle is significantly detectable, a process we detail in the next section.


\textbf{Calibrating Cosmic Expansion---} We have demonstrated that the 21-cm acoustic signal is not a single standard ruler, but the superposition of BAOs and VAOs as two physically distinct oscillatory components. Their relative prominences wax and wane over time and scale, and their phases need not align. Recently, Ref.~\citep{Montefalcone:2025mbg} showed that while BAOs trace oscillations in the baryon density field $\delta_b$, relative velocities trace $\eta \sim \delta_b^2$, producing a characteristic phase shift between the ``wiggle''-only BAO and VAO templates in Fig.~\ref{fig:FIG1} (see \citep{Green:2020fjb, hahn21, ferraro12} for related second-order effects.) Having established the detectability of and distinguishability between BAOs and VAOs, we now inquire in this section how modeling them improperly may erroneously bias $H(z)$ measurements.

\begin{figure}[t]
    \centering
    \includegraphics[width=0.5\textwidth]{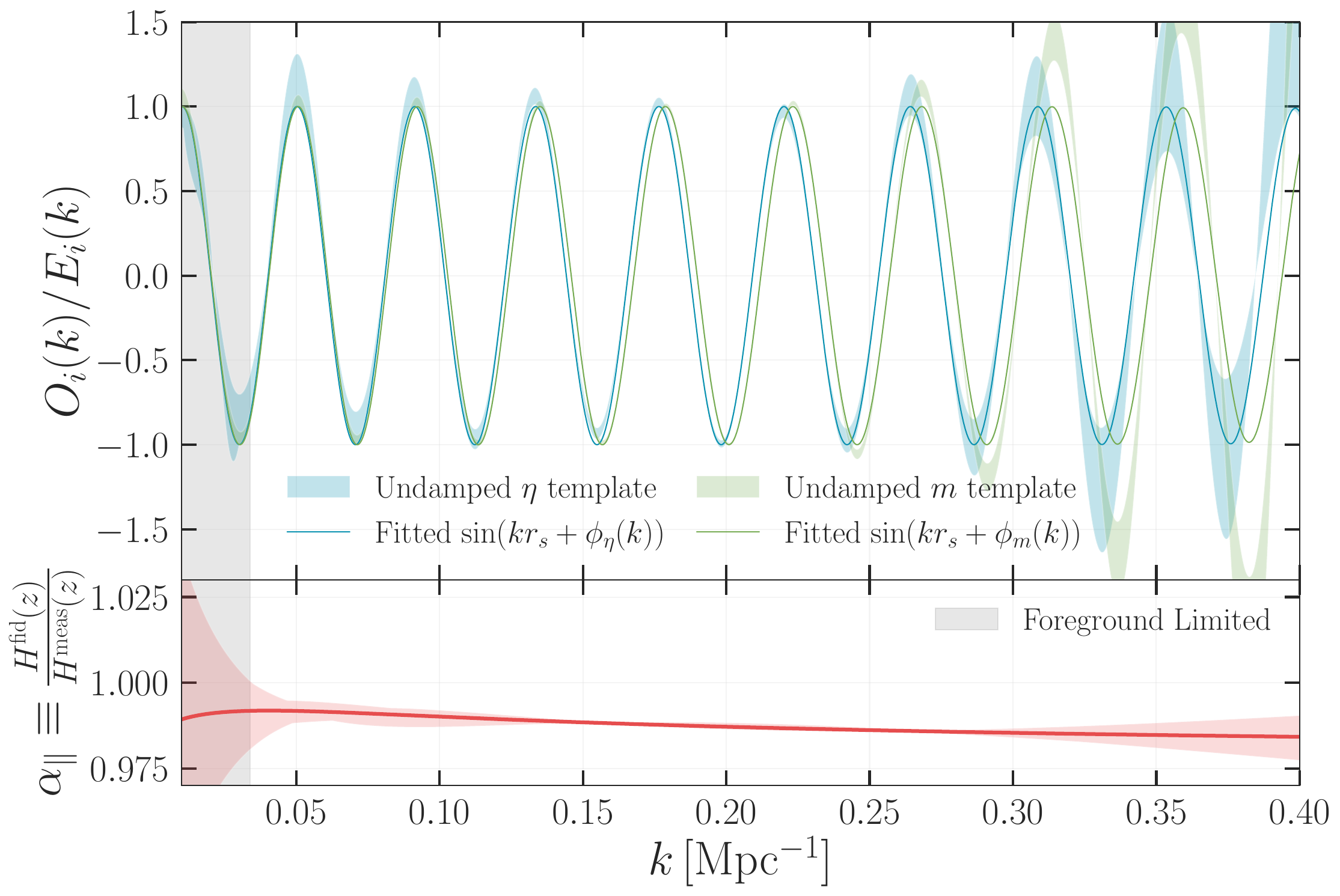}
    \caption{\textbf{(Top):} Undamped oscillatory component $\widetilde{O}_i(k)$ of $P_m(k)$ (green) and $P_\eta(k)$ (blue) and the fitted sinusoid (solid). Fitting three envelope functions $E_i(k)$ to $O_i(k)$ peaks, troughs, and both extrema, we show their differences as uncertain shaded regions. \textbf{(Bottom):} The incorrect AP parameter estimate from modeling a VAO-only wiggle as a distorted BAO-only template. If assuming a matter-only model, the inferred $H(z)$ (red) deviates from truth by $\sim 2\%$ across the majority of observable modes.}
  \label{fig:FIG3}
\end{figure}

If the total $\Delta^2_{21}(k,z)$ is a superposition of BAO and VAO components, the maximum phase difference between them is set by the phase offset between $P_m(k)$ and $P_\eta(k)$. We measure this by first isolating the normalized ``wiggle''-only templates $O_i(k), \, i \in (m, \eta)$ using the baseline extraction procedure described in the Supplementary Material. Then, we undamp the wiggles by dividing out a fitted envelope function $E_i(k)$, yielding a wave $\widetilde{O}_i(k)$ that undulates between $-1$ and 1. The top panel of Fig.~\ref{fig:FIG3} compares these matter and velocity ``wiggle'' templates, with shaded regions capturing the uncertainty in the undamping procedure by fitting $E_i(k)$ to the $O_i(k)$ peaks, troughs, and both extrema. Consistent with Ref.\citep{Montefalcone:2025mbg}, the undamped matter and velocity template nearly coincide at low $k$ and grows increasingly shifted toward higher $k$.  We find a smaller peak shift than in previous semi-numerical studies \citep{munoz19, Sarkar:2022mdz}, which we attribute to our use of $P_\eta(k)$ at $z=50$ instead of the standard $z_\mathrm{kin} \approx 1060$. The latter retains non-negligible decaying velocity modes present near kinematic decoupling but becomes increasingly suppressed toward cosmic dawn. Our methods are consistent with recent 21-cm treatments that use scale-dependent growth factors to track the differing evolution of density and velocity perturbations \citep{flitter25}.

These phase differences impact 21-cm derived measurements of $H(z)$. Should future 21-cm analyses neglect velocity fluctuations and expect that only matter fluctuations contribute 21-cm wiggles, or vice versa, one might subsume these phase differences into an improper estimate of distortions along the line of sight \citep{thelie25}. Explicitly, one can find how maximally incorrect this assumption is by treating the $\eta$ phase template as a stretched matter template via
\begin{equation}
    \sin(k r_s + \phi_\eta(k)) \sim \sin\left(\frac{k r_s}{\alpha} + \phi_m(k)\right), 
\end{equation}
where $\alpha$ is the Alcock-Paczy\'nski (AP) parameter. As 21-cm interferometers are mostly sensitive to modes along the line of sight, we express $\alpha_\parallel = H^\mathrm{fid}r^\mathrm{fid}_s/(H^\mathrm{meas} r^\mathrm{meas}_s)$ as a ratio of fiducial and measured quantities. Assuming the sound horizon is externally constrained, the bottom panel of Fig.~\ref{fig:FIG3} shows the relative error between $H(z)$ measurements over scale. We find measurements of $H(z)$ may be off by $\sim 2\%$ across the majority of observable $k$-modes, so a percent-level measurement of $H(z)$ at cosmic dawn must properly model BAOs and VAOs. 



\textbf{Discussion---} In this work, we have studied for the first time how the 21-cm power spectrum during cosmic dawn has distinct BAO and VAO components, demonstrated how to separate them, and touched upon the cosmological implications. Though the amplitude of either acoustic feature on the 21-cm signal is controlled by astrophysics, the positions of their peaks and troughs are cosmologically frozen-in. We have demonstrated that the amplitudes of BAO and VAO peaks are strong functions of scale, and do not grow or shrink monotonically over redshift. Rather, we find that VAOs dominate at early times yet briefly vanish at the transition between Lyman-$\alpha$ coupling and X-ray heating. BAOs dominate the signal at later times, especially at higher-$k$ (which are more readily measurable). 

Given this BAO-VAO decomposition, we then forecast how well upcoming experiments can detect these wiggles and constrain the expansion rate. We find that though ``moderate'' foreground removal scenarios can achieve a $\sim 2 \sigma$ detection of the total acoustic bumps by SKA at  $z\gtrsim 10$, ``optimistic'' mitigation scenarios are necessary to separate the contributions from matter and velocity. Our goal here is not to perform a full joint cosmological and astrophysical parameter inference. Rather, we ask whether the BAO and VAO template components are, in principle, detectable and distinguishable, assuming the fiducial astrophysical model. We argue that joint modeling of BAOs and VAOs is critical at cosmic dawn, as using either template alone produces a scale-dependent error in $H(z)$ constraints of $\sim 2 \%$. Encouragingly, using \textit{SKA} and/or extending our 21-cm models to reionization \citep{thelie26, libanore25} will yield more significant detections.

Several caveats may impact a future detection of acoustic oscillations in the 21-cm spectrum. VAOs arise as feedback on early galaxies, and future work is needed to better constrain the size of this effect, as it remains a large source of uncertainty~\cite{schauer21}. Other radiative feedback effects \citep{barkana99, mesinger08, okamoto08} and novel constraints on high-redshift ultra-low metallicity SFGs \citep{venditti25} may impact overall VAO amplitudes. Bursty star formation as suggested by \textit{JWST}-derived UVLFs \citep{mason23, mirocha23, shen23, munoz26} inevitably breaks the galaxy-halo connection assumed here, though it may effectively add Poisson fluctuations on the 21-cm field that elevates $\Delta^2_{21}(k,z)$ into a more detectable realm \citep{Libanore:2023oxf}. Conversely, purely adiabatic fluctuations in a strict-$\mathrm{\Lambda}$CDM cosmology will naturally induce residual baryon-CDM compensated fluctuations from baryon-photon decoupling, which may reduce local baryon fractions and star formation rates up to tens of percent \citep{jessop26} which may lower high-$z$ BAO sensitivities. 

In summary, a future detection of 21-cm fluctuations will provide powerful insights on galaxy evolution and cosmology at previously inaccessible redshifts. But to harness its full constraining power, one must model the full acoustic structure of 21-cm fluctuations as the sum of its matter and velocity-induced components. As they trace different physical processes, modeling how BAOs and VAOs grow and diminish, how they sum together to be observable, and the conditions in which they become separable is imperative to separate astrophysical effects from cosmological information. The decomposition introduced in this work is a key step toward preparing the theory needed for upcoming observations, and thus toward using the 21-cm signal as a probe of structure formation and cosmic expansion during the first billion years of cosmic history.

\begin{acknowledgments}
We would like to thank Marc Kamionkowski, Benjamin Wallisch, Yacine Ali-Ha\"imoud, and Anthony Pullen for helpful discussions. HAC was supported by the National Science Foundation Graduate Research Fellowship under Grant No.\ DGE2139757, and by the Simons Society of Fellows through Grant No. SFI-MPS-SFJ-00010459. GM acknowledges support by the Writing Fellowship of
the Graduate School of the College of Natural Sciences at
the University of Texas at Austin. JBM was supported by NSF Grants AST-2307354 and AST2408637, and NASA through grant JWST-GO-03224.   EDK acknowledges support from the U.S.--Israel
Binational Science Foundation (NSF-BSF grant 2022743 and BSF grant
2024193) and the Israel National Science Foundation (ISF grant 3135/25),
as well as from the joint Israel--China program (ISF--NSFC grant
3156/23). AV acknowledges funding from the Cosmic Frontier
Center and the University of Texas at Austin’s College of Natural Sciences. We acknowledge the use of \texttt{CLASS}~\cite{Blas:2011rf},  \texttt{zeus21}~\cite{munoz23,munoz23b,Cruz:2024fsv}, and the Python packages \texttt{Matplotlib}~\cite{Hunter:2007mat}, \texttt{NumPy}~\cite{Harris:2020xlr} and~\texttt{SciPy}~\cite{Virtanen:2019joe}.
\end{acknowledgments}

\bibliography{refs}
\clearpage
\onecolumngrid
\begin{center}
  \textbf{\large Supplementary Material for The Rise and Fall of Acoustic Oscillations at Cosmic Dawn}\\[.2cm]
  \vspace{0.05in}
  {Hector Afonso G. Cruz, Gabriele Montefalcone, Julian B. Mu\~noz, \\ Ely D. Kovetz, and Alessandra Venditti}
\end{center}
	\twocolumngrid

\section{Stellar Population Modeling}\label{sec:AppSFE}
Here we detail how matter overdensities and relative velocities modulate different stellar populations. More overdense regions yield relatively more haloes which boost the SFRD identically across populations, which we capture as a $\delta$-augmented halo mass function (see \citep{munoz23, Cruz:2024fsv} for the full prescription). However, population II and III stars differ in their star-forming efficiency (SFE). We assume population II SFGs form in atomic-cooling haloes (pre-enriched by previous Pop III stars) of mass $M_h \gtrsim 10^8 \, M_\odot$\citep{oh03}, massive enough to be modeled as impervious to relative velocities. Conventionally, its SFE is a double power-law in halo mass \citep{moster13, furlanetto17, sabti22b} described by
\begin{equation}
    \mathrm{SFE}^\mathrm{II} = \frac{2 \epsilon_*^\mathrm{II} \exp \left( -M_\mathrm{atom}/M_h\right)}{\left(M_h / M_p^\mathrm{II}\right)^{\alpha_*^\mathrm{II}}+\left(M_h / M_p^\mathrm{II}\right)^{\beta_*^\mathrm{II}}} ,
\end{equation}
with an extra exponential duty cycle in terms of the atomic mass turnover threshold $M_\mathrm{atom}$ below which cooling becomes inefficient. All the above quantities use the best-fit parameters from \textit{HST}+\textit{JWST} UV Luminosity Function (UVLF) data in the rightmost column of Table E1 in Ref.~\citep{munoz23b}. In contrast, population III stars in molecular-cooling haloes $(10^6 \lesssim M_h \lesssim 10^8 \, M_\odot)$ exhibit a SFE suppressed by the $\eta(\textbf{x})$ field. We parameterize this by raising the critical halo mass threshold to form Pop III stars efficiently, via $M_\mathrm{mol} \propto M_\mathrm{mol,0}f_{\eta}$ \citep{Maio:2010qi, stacy11, greif11, naoz13} with the feedback parameter $f_{\eta}$ as computed in Ref.~\citep{munoz22}. The Pop III SFE in a halo of mass $M_h$ is modeled as flat (alternatively, $\alpha_* = \beta_* = 0$) with exponential dependence on the mass threshold via
\begin{equation}
    \mathrm{SFE}^\mathrm{III} = \epsilon_*^\mathrm{III} \exp(-M_\mathrm{mol}/M_h)\exp(-M_h /M_\mathrm{atom}).
\end{equation}
Using the procedure in Ref.~\citep{venditti25}, we perform a best-fit to the UVLFs from the AMORE6 Pop III galaxy candidate \citep{morishita25} and the upper limits on the Pop III UVLF in Ref.~\citep{fujimoto25}. Assuming a UV boost factor of $\sim 3$ \citep{inayoshi22} (but could plausibly be as high as $\sim 6$ \citep{liu25}) yields an SFE amplitude $\epsilon_*^\mathrm{III} = 10^{-3.05}$ and a redshift-independent upper threshold $M_\mathrm{atom} = 10^{8.39} \, M_\odot$. All other unspecified parameters are fixed to defaults in the first two columns of Table I in Ref.~\citep{Cruz:2024fsv}.

\section{21-cm Formalism Modifications}\label{sec:App21cmMod}

We note an important modification to $\bar{T}_{21}$ that will allow an end-to-end separation of baryons from dark matter. Since $\bar{T}_{21}$ tracks IGM gas fluctuations, the brightness temperature must be proportional to the fractional baryonic density $\delta_b$ \citep{Furlanetto:2006jb, Pritchard:2008da, Pritchard:2010pa, pritchard12}, in contrast to previous works that instead use the total matter density $\delta$ \citep{madau96, barkana06, Barkana:2016nyr}. Baryonic and total matter perturbations are approximately adiabatic up to a few percent through cosmic dawn. Indeed, this approach is used in semi-numerical codes \citep{mesinger11, Cohen:2016jbh} where tracking separate boxes for $\delta_b$ and $\delta_\mathrm{CDM}$ becomes computationally cumbersome. Our BAO/VAO separation implements a $\delta \rightarrow \delta_b$ substitution in the $T_{21}$ prefactor to mitigate these percent-level differences. This change yields a negligible difference on acoustic amplitudes at low $z$ and $\sim 25\%$ larger acoustic amplitudes at high $z$. We do not modify $\delta$ anywhere else in our formalism \citep{Cruz:2024fsv}, as other quantities like the halo mass function or star formation rate density depend on total matter density. The substitution lowers BAO amplitudes, but will still dominate over VAOs toward later redshifts as accreting haloes grow increasingly immune to the velocity differential. We find that this change does not impact the BAO/VAO hierarchy or the phase extraction procedure beyond the sub-percent level.

For self-consistency, the redshift at which we use velocity transfer functions is also modified from $z_\mathrm{kin}$ to $z=50$. Recently, Ref.~\citep{Cruz:2024fsv} showed that VAOs in the 21-cm power spectrum are simply exponential correlations of the scalar power spectrum of the relative velocity $P_\eta(k)$ itself. To match existing semi-numerical codes \citep{munoz22}, the acoustic peaks were assumed to have been fixed at kinematic decoupling and decayed strictly as $\propto a^{-2}$. But since decaying modes present at $z_\mathrm{kin}$ only become negligible toward the end of the dark ages, the shape of $P_\eta(k,z)$ is only roughly independent of redshift for $z \lesssim 100$ \citep{ferraro12, Yoo:2011tq}. Thus, we find that accurate models of velocity-modulated signatures in the IGM must use spectral templates anchored at $z \approx 50$, where baryons have largely relaxed. Fortunately, hydrodynamical IGM simulations show that relative velocities raise the $M_\mathrm{mol}$ threshold independently of redshift as a simple power-law scheme \citep{schauer19, kulkarni21, munoz22}, and should impact star-forming duty cycles separably as currently implemented in {\tt\string Zeus21}.

\section{Baseline \& Phase Extraction}\label{sec:AppPhase}

\begin{figure}[t]
    \centering
    \includegraphics[width=0.45\textwidth]{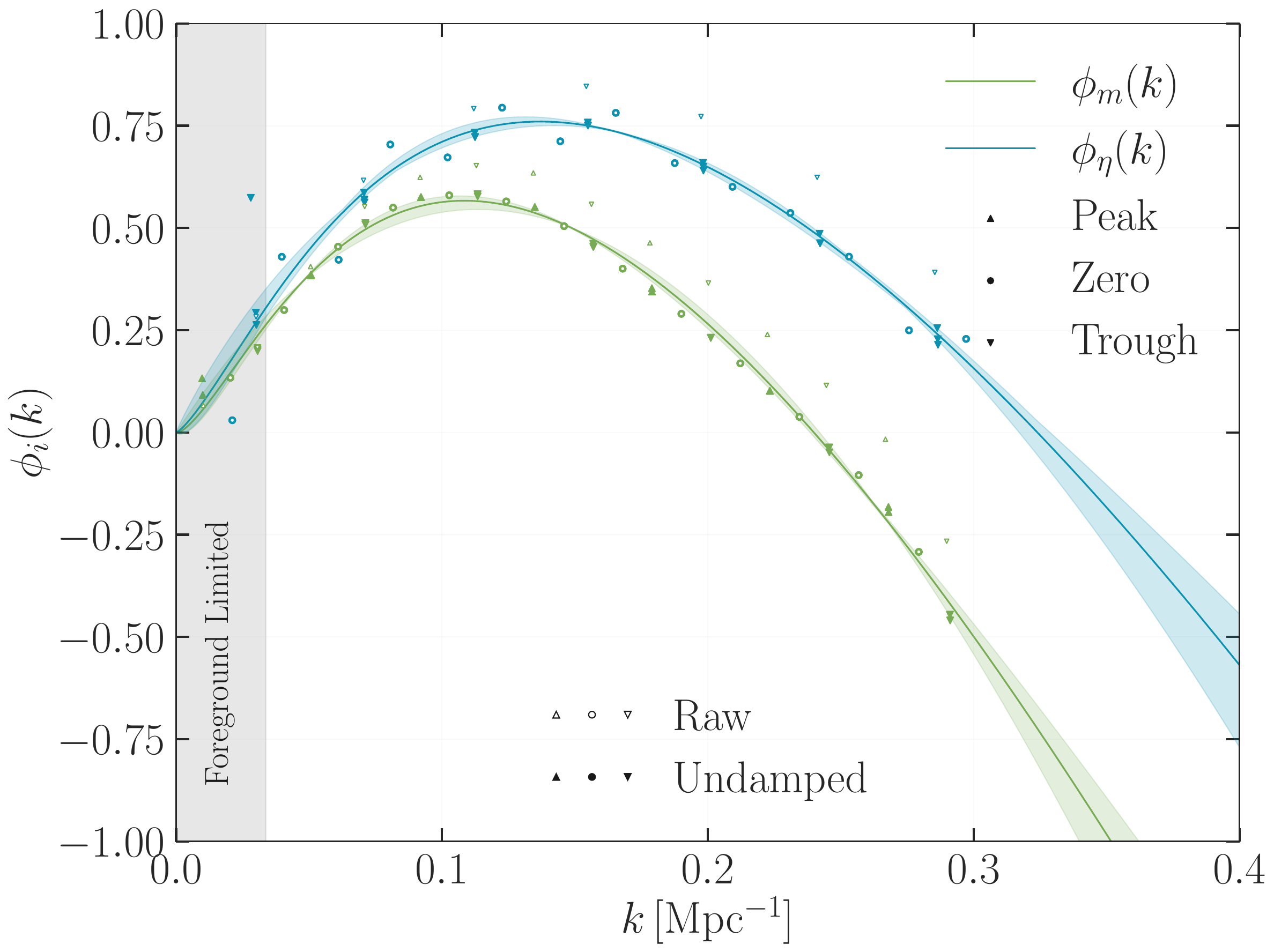}
    \caption{The extracted scale-dependent phase for wiggles in $P_m(k)$ (green) and $P_\eta(k)$ (blue). Shaded areas capture the different estimates of $\phi_i(k)$ when using different envelope functions to undamp the normalized wiggle templates $O_i(k)/E_i(k)$. Hollow markers capture the phase shift in raw $O_i(k)$ extrema and zeroes, while solid markers capture the phase shift in the undamped sinusoids as in Fig.~\ref{fig:FIG3}.}
  \label{fig:FIGAPP}
\end{figure}

In this section, we describe our baseline extraction procedure that lets us isolate the ``wiggle'' and ``no-wiggle'' components in the 21-cm power spectrum. The cumulative influence of scale-dependent physical processes introduce smooth curvature over a wide range of $k$-modes in the matter and relative velocity transfer functions, meaning their baselines are not ideal broken power-laws to which we can fit. Thus, fitting for the zeros on top of which either waveforms oscillate is difficult in Fourier space.  It is easiest to perform a discrete inverse sine-transform on $P_m(k)$ and $P_\eta(k)$, interpolate over the BAO and VAO bumps in configuration space, and re-transform back into $k$-space to get $P^\mathrm{nw}_m(k)$ and $P^\mathrm{nw}_\eta(k)$ in accordance with App. C.1 of \citep*[see also \citep{Hamann:2010pw}]{Baumann:2017gkg}. We alternate these as inputs into {\tt\string Zeus21} to yield BAO, VAO, and ``no-wiggle'' 21-cm power spectra. The results are the normalized ``wiggle''-only templates $O_{21,i}(k) \equiv P_{21,i}(k)/P^\mathrm{nw}_{21,i}(k) - 1, \ i \in \{\mathrm{BAO}, \mathrm{VAO}\}$ in Fig.~\ref{fig:FIG1}. We find that our measured BAO and VAO amplitudes do not change beyond the sub-percent level with different bump interpolation methods, Fourier inversion schemes, or other differences in baseline definitions.

In the penultimate section of the main text, we outline how these ``wiggle''-only templates can be used to estimate incorrect $H_0$ constraints should relative velocities be neglected. We found that the maximum phase difference between BAOs and VAOs across all epochs is merely the phase difference in the raw matter and scalar relative velocity power spectra $P_m(k)$ and $P_\eta(k)$. Our inverse sine transform procedure allows us to isolate the normalized ``wiggle''-only templates $O_j(k) = P_j(k)/P_j^\mathrm{nw}(k) - 1, \ j \in \{m, \eta\}$ as similar damped sinusoid-like functions. Similar to previous prescriptions \citep{Baumann:2017gkg}, we parameterize them with a damping envelope
\begin{equation}
    E(k) = (1-a\exp(-bk^c))d\exp(-fk^g)
\end{equation}
and a sinusoid with a scale-dependent phase, via the functional fit
\begin{align}
    O_i(k) &= E_i(k) \sin \left( kr_s + \phi_i(k)\right), \label{eq:OiK1}\\
    \phi_i(k) &= \frac{\phi_{\infty,i}}{1+\left(k_{*,i}/k\right)^{\xi_i}} + \phi_{2,i}k \label{eq:OiK2}
\end{align}
We treat the envelope parameters $(a, b, c, d, f, g)$ and phase parameters $(\phi_{\infty,i}, k_{*,i}, \xi_i, \phi_{2,i})$ as nuisance quantities fitted with a nonlinear least-squares approach. The phase function mirrors those used in BAO analyses \citep{Baumann:2015rya, Baumann:2017gkg, Baumann:2017lmt, Baumann:2019keh, Green:2020fjb}, albeit with an extra linear nuisance parameter to capture extraneous physics \citep{Green:2020fjb}. We note that the raw ``wiggle'' spectra are not perfectly sinusoidal. The $O_i(k)$ are the result of the total impact of scale-dependent shifts and nonlinear skew from pre-Recombination interactions between baryons and light relics, radiation driving, baryon loading, and Silk damping \citep*[see also \citep{Pan:2016zla, Baumann:2015rya}]{Baumann:2017lmt, Baumann:2017gkg, Montefalcone:2025mbg, Montefalcone:2025unv}. Hence, the $O_i(k)$ extrema are not uniformly shifted, and thus the peaks are fit by slightly different envelope and phase parameters than the troughs.

We find that our baseline extraction process does not impact our phase estimate in our undamped $\widetilde{O}_i(k)$ beyond a few tenths of a percent. In Fig.~\ref{fig:FIGAPP}, we show the effect of our undamping procedure on the extracted phase templates $\phi_i(k)$. The raw $O_i(k)$ peaks and troughs almost always overestimate the phase when compared to the undamped $O_i(k)/E_i(k)$ templates in both the matter and relative velocity wiggles. Expectedly, dividing out the envelope function does not impact the phase extracted from the zeroes. The error in phase extraction is given by shaded regions around the $\phi_i(k)$ functional fits, and are computed by the span of envelope functions fitted to the peaks, troughs, and both extrema. These three estimates of $E_i(k)$ yield three phase estimates for each undamped extremum or zero. Our plot shows that, unless we are concerned about measuring $O_i(k)$ features in the $k$-realm limited by foregrounds, the three phase estimates for each undamped extremum vary by less than the percent level. One could argue that improper baseline subtraction $P_i^\mathrm{nw}(k)$ may bias our phase extraction procedure. Yet since the undamped extrema vary so little, we find that dividing out our three estimates for $E_i(k)$ yields a sufficiently modeled $\phi_i(k)$. 

\end{document}